\documentclass[onecolumn,secnumarabic,amssymb, nobibnotes, showkeys, aps, prd]{revtex4-2}
\usepackage{graphicx} 

\usepackage{hyperref}
\usepackage{amsmath}
\usepackage{amssymb}
\newcommand{\env}[1]{\texttt{#1}}

\begin{document}
\title{MCNP-GO: A python package for assembling MCNP input files with a systems engineering approach}
\author{A. Friou}
\date{\today}
\email{alexandre.friou@cea.fr}
\affiliation{CEA, DAM, DIF, F-91297 Arpajon, France}

\begin{abstract}
This article introduces MCNP-GO (\href{https://github.com/afriou/mcnpgo}{https://github.com/afriou/mcnpgo}), a Python package designed to manipulate and assemble MCNP input files, allowing users to assemble a set of independent objects, each described by a valid MCNP file, into a single cohesive file. This tool is particularly useful for applications where precise modeling and positioning of equipment are crucial. The package addresses the challenges of managing large databases of MCNP input files, ensuring reliability and traceability through configuration management systems. MCNP-GO provides functionalities such as renumbering, extracting subsets of files, transforming files, and assembling files while managing collisions and materials. It also keeps track of the operations performed on files, enhancing traceability and ease of modification. The article demonstrates the package's capabilities through a practical example of assembling an MCNP input file for a tomographic experiment, highlighting its efficiency and user-friendliness. MCNP-GO is designed for users with minimal Python knowledge.
\end{abstract}


\maketitle

\section{Introduction and motivation}

This article introduces MCNP-GO (\href{https://github.com/afriou/mcnpgo}{https://github.com/afriou/mcnpgo}), a Python package for manipulating and assembling MCNP input files \cite{MCNPREF}. This package allows users to assemble a set of independent objects (i.e. equipment, experiment room, detector, etc ...) in a single file. The nature of the object to be assembled is not important, but each object should be described by an independent and valid MCNP file. To better grasp the reasoning and ideas behind MCNP-GO, it is important to give some context. 

First, consider the needs of a person responsible for radiation protection. There are possibly many facilities to deal with, each are equipped with dosimeters to monitor radiations and each possess their own source of radiations and various equipment. It is necessary to properly model those facilities to compute dose maps, which could be done with MCNP. Dealing with such a database of MCNP input files is cumbersome, but must be done reliably, ideally with a configuration management system, considering the legal and health risks that workers could face in case of an error. Moreover, modifying such database to add or update equipment or sources is a time intensive and error prone task. The sum of it could then become a real burden to deal with, making it hard to respond quickly and reliably. In this situation, a tool to assemble MCNP input files would be of great help.

Radiation protection usually does not require to precisely model where the equipment is. In most situations, a precision of the order of a cm is sufficient to perform dose calculations. However, that could be crucial for a tomographic experiment, where an X-ray source is used to radiograph an object of interest in order to precisely infer the object density map and materials. To that end, the position and orientation of the source, of the object of interest and of the detector, should be known with great precision. Otherwise, position uncertainties could hinder tomographic reconstruction. Here, one would need a tool to position MCNP input files relative to each other with ease.

The package presented in this article is a response to these concerns. It was originally designed to address the following problem: "How to generate on demand an MCNP geometry of a fully equipped facility?". Knowing that the experimental set-up is likely to evolve on a day-to-day basis, we need a tool that can quickly generate an up-to-date input file. It should be as close as possible to the reality in terms of position and orientation, but each object is modeled in MCNP with an approximation of its geometry which is deemed sufficient for the problem at end. We thus have a database of objects, from which the user will pick to assemble his input file. The design step of an object from drawings or CAD files is done upstream, and does not concern this package. For this step, the user may refer to a dedicated software or package such as SuperMC, or GEOUNED \cite{WU2009,GEOUNED}.

Ideally, each object in the database would be managed using configuration management system such as Git or SVN. This would involve tracking and controlling changes made to the object over time, using a configuration management system. This can be especially important in situations where the object is being used in a critical or sensitive context, where it is important to ensure that it is being used correctly and consistently. One of the main advantages of the approach presented here, is that the final MCNP file assembled by the package keeps tracks of the assembling process. In practice, objects versions numbers and applied transforms are detailed in the header file.

To better understand the reason for this package and how it works, it is necessary to explain the problems encountered when trying to assemble two MCNP input files. An input file is composed of cards describing cells (i.e., physical volumes of matter), surfaces, transformations and materials. Each card is identified by a unique number. As a result, it is generally not possible to join two files together without conflicts between data cards or conflicts with overlapping cells.

Moreover, we would also need to place object 1 of file 1 relative to object 2 of file 2, by modifying transformation cards. The latter allow to apply a translation and a rotation to a surface or a cell. They are therefore a way to position objects relative to each other. In the general case, this turns out to be far from simple, especially if this operation had to be done manually.

The operations become even more complex when the input file contains several objects. However, if these two objects are independent, then we can in principle create two independent files each containing one object. This is one of the basic principles of the package, it works with independent objects. To facilitate the task of building a database of objects, the package has an extract capability, which allows to cut out a part of a file to make another independent file. This way, we can re-process part of an old file and reuse it.

The answer to all these problems, the guiding philosophy, is to consider each MCNP file as an object in the sense of object-oriented programming. The following operations are possible on this object:
\begin{itemize}
    \item Transformation (translation and rotation).
    \item Re-numbering of surfaces, cells and transformations.
    \item Extraction of a subset of the object to generate a new object.
    \item Insertion of an object into another (managing collisions and materials).
\end{itemize}
In this paper we will see with the help of examples how MCNP-GO works in practice. It was conceived for people with no knowledge of Python and is in daily use at CEA to build input decks with a few thousands cells. It was fully written in Python and only has a few dependencies (NumPy, json) to keep installation as simple as possible. 

There are a number of python packages that were developed over the years to facilitate MCNP input file handling (see \cite{MontePy} for a review). However, in order to answer the issues stated above, MCNP-GO needs two crucial features: file assembly and file transformation. The first feature could be done with MontePy \cite{MontePy} or numjuggler \cite{Numjuggler} as they both possess re-numbering capabilities. Note that numjuggler has an option to assemble two files providing that the proper renumbering was done before. In principle, one could harness theses packages capabilities to write a program that automate file assembly. However, the second feature, file transformation, to the best of the author knowledge, is not supported by any package. Each of them was designed with a use case in mind and their code architecture reflects it. MCNP-GO answers to other issues stated above, which to the author point of view is shared by many MCNP day-to-day users, and was not addressed before. 
	
This paper is organized as follows. Section \ref{Requirements and caveats} describes the requirements in terms of cell block structure necessary to assemble two files. The author believes that while it is not a heavy burden, it needs to be stated clearly and put forward. We then review the main package features in section \ref{Package features}, with the help of conceptual examples. We do not to delve into too much details but rather try to show the underlying logic of the package. Finally, a practical example is given in section \ref{Example use case} where we build an MCNP input file corresponding to a tomographic experiment. One could jump to this section to get a general idea of the package's capabilities and then come back to the previous section for details.

\section{Requirements and caveats}
\label{Requirements and caveats}

In order to be able to assemble two files together, i.e. to insert one file in an other, the cell block cards must possess a particular structure. Consider the following cell block cards:
\begin{verbatim}
1 83 -7.13 -1                         $ scintillator
2 13 -2.7  -2 3                       $ scintillator cover
3 13 -2.7  -4 5 1                     $ detector box
4 26 -7.9  -6                         $ base steel plate
6 14 -2.4  -9                         $ mirror
10 100 -1.205e-3 (1 #3 9) (-3:-4)     $ air
11 0 (6 4 2)                          $ graveyard
\end{verbatim}
The last two cells, 10 and 11, are remarkable:
\begin{itemize}
    \item The second-to-last cell describes the ambient medium (here, air) in which bathe the other cells, and this is the only cell like this. 
    \item The last cell describes the external world (where particles are killed). In this cell, surfaces 6, 4 and 2, define the bounding surface, delimiting the external world from the interior.
\end{itemize}   
The only prerequisite required by the tool is to place the cell describing the ambient medium in the second-to-last position, and that describing the external world in the last position. This structuring is generally very easy to achieve. Note that this is necessary only for the package insert feature (see section \ref{Assembling files}). The other features do not require any particular structuring of the file. As we shall see, inserting a file into another is based in part on the boundary surface of the object.

In the rest of the article we will refer to the second-to-last cell as "gas cell", and the last cell as "graveyard cell".

The function of MCNP-GO that interprets MCNP files is not as efficient as the one implemented in the MCNP code. While it could be perfected with more developments, a few caveats still exists:
\begin{itemize}
    \item A card description should not be interrupted by a comment line (starting with "c"). Rather, comment using the dollar symbol at the end of the line.
    \item Vertical input format is not implemented.
    \item Universes and fill cards must be defined in the cell block.
\end{itemize}
MCNP-GO tries to catch and correct the input file and issues warnings, but exceptions might still occur. In particular, the output file is formatted as little as possible, in order to facilitate comparison between input files and the final (assembled) output file.	

\section{Package features}
\label{Package features}

In this section, we do a quick review of the package's main functions. We do not delve into too much details, but rather try to highlight how the package could be used to simplify work flow. The main functions, dealing with assembly and transformation of files are dealt at the end of the section.

\subsection{Renumber}
This is the first and simplest of functions. In practice it is not needed since files are automatically renumbered, if necessary, during assembly. However, it could be useful for readability purposes. The following code renumbers cells from 1, surfaces from 10 and transformations from 100:
\begin{verbatim}
    from mcnpgo.mcnpgo import *
    obj1 = go("my_file.mcnp") # Read file
    obj1.Renum(cell=1, surf=10, trans=100) # Renumber
    obj1.WriteMCNPFile("file_renum.mcnp")  # Save file
\end{verbatim}
It is not possible to renumber material cards with this function. But they are renumbered (if necessary) automatically during assembly. More details are given about material cards in section \ref{Assembling files}.
	
\subsection{Re-process legacy files}
\label{Re-process legacy files}
We describe here the extract function, which is very useful for re-processing legacy files, by retrieving only a part of them (see section \ref{Example use case} for an example use case. Consider the following code, that extract cells from 80 to 90 of the input file:
\begin{verbatim}
    from mcnpgo.mcnpgo import *
    obj1 = go("my_file.mcnp") # Read file
    new_obj = obj1.Extract(range(80,91)) # Extract cells from 80 to 90
    new_obj.WriteMCNPFile("file_extract.mcnp")  # Save file
\end{verbatim}
The function proceeds recursively to retrieve all the surfaces, cells, transformations and materials that depend on the list of cells provided. However, it does not try to guess what could be the bounding surface of the new object, which could prove difficult. Instead, the new object created is placed by default in a 20m radius sphere, in order to encompass all the cells. This is generally sufficient for the majority of objects. It is then up to the user to rework the extracted object if he wishes to give it a more suitable bounding surface.
	
Note that the extract function is useful to cast a file with the structure described in section \ref{Requirements and caveats}, necessary for assembling two files.

\subsection{The problem of cell transforms cards}

A well-known problem for MCNP users concerns the use of cell transformation, related to \texttt{trcl} cards, on surfaces with a number greater than 999. In these cases, a fatal error is generated and it is impossible to run the code. It is then generally necessary to circumvent the problem by duplicating cells or renumbering affected cells or surfaces, a cumbersome and error prone procedure if done manually. However, this operation can be done with the following code:
\begin{verbatim}
    ... # You have assembled your input file and reached this point
    obj.ResolveTRCL() # Renumber cells or surfaces that would pose problem with transforms cards
    obj.WriteMCNPFile("result.mcnp") # Save file
\end{verbatim}
The \env{ResolveTRCL} function performs the renumbering of affected surfaces or cells. This procedure is not applied systematically in order to speed up processing. However, the user can call it if a problem arises.
	
\subsection{Assembling files}
\label{Assembling files}
    
In this section we discuss two methods for assembling files and dealing with collisions. Other possible methods are discussed in section \ref{Conclusions and future work}.
	
\subsubsection{Assembling files using bounding surfaces}

\begin{figure}
\includegraphics[scale=0.5]{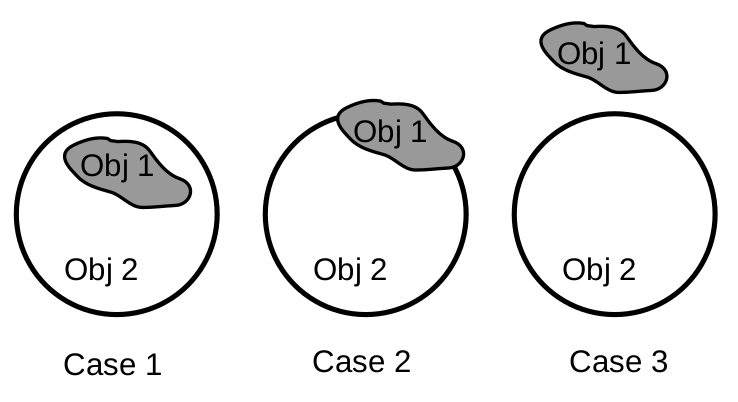} 
\caption{Cases covered by the \env{Insert} function. \label{cases}}
\end{figure}

Assembling two files can be done with the following code:
\begin{verbatim}
    # Load MCNP files
    obj1 = go("object1.mcnp")
    obj2 = go("object2.mcnp")
    
    # Assemble two files
    obj2.Insert(obj1)
\end{verbatim}
In this example, the \env{Insert} function inserts \env{obj1} in \env{obj2}, so that it now contains entirely \env{obj1}.  In practice, the bounding surface of \env{obj1} is added to the description of the last two cells of \env{obj2}, i.e, the gas and graveyard cells (see section \ref{Requirements and caveats}). This is therefore an insertion by "bounding surface". It means that \env{obj1} can be anywhere in the gas cell of \env{obj2}, or outside in the graveyard, without collision issues, as illustrated in Figure \ref{cases}. However, if the bounding surface of \env{obj1} meets a cell of \env{obj2}, then a collision problem arises. It would be possible to make \env{obj1} bounding surface prioritized over the cells of \env{obj2}, but it would be sub-optimal to allow it.

During assembly, cell, surface and transformation cards are renumbered to avoid any conflict and the material cards are merged. If a material card of \env{obj1} is already present in the list of materials of \env{obj2}, then it is not added. Otherwise, a new material card number is found (if necessary) and the material is added to the list. Material cards are compared scrupulously, i.e, cross-section libraries, proportions and elements must be identical for two material cards to be considered identical. Note that corresponding material cards \texttt{mpn}, \texttt{mx} and \texttt{mt}, are carried if present.

By default, insertion covers all cases illustrated in Figure \ref{cases}. Two other options are useful when the user knows that object 1 will always be completely contained within object 2 (case 1) or completely outside of object 2 (case 3). These situations are specified during insertion by the keywords \env{location="inside"} and \env{location="outside"}:
\begin{verbatim}
    # obj1 is completely contained within obj2 (case 1)
    obj2.Insert(obj1, location="inside")
    
    # obj1 is completely outside of obj2 (case 3)
    obj2.Insert(obj1, location="outside")
\end{verbatim}
In practice, this allows to optimize cells description:
\begin{itemize}
    \item In case 1: the bounding surface of object 1 is added to the gas cell of object 2, but not to the graveyard cell.
    \item In case 3: the bounding surface of object 1 is added to the graveyard cell of object 2, but not to the gas cell.
\end{itemize}	
Note that the keyword \texttt{location} is optional, the default mode will always work.

\subsubsection{Assembling files with cells exclusion}
We saw how we could assemble two files by copying the bounding surface of the inserted object to the other. An other way of assembling files is to exclude the cells of the inserted file from the gas cell of the other using MCNP exclusion operator \verb!#!. This can be done with the following code:
\begin{verbatim}
    # Load MCNP files
    obj1 = go("object1.mcnp")
    obj2 = go("object2.mcnp")
    
    # Assemble two files
    obj2.InsertCells(obj1)
\end{verbatim}
In this process, the last two cells of the inserted object, \texttt{obj1}, are deleted.
This way of assembling files is certainly less efficient in terms of computer time. Indeed, \texttt{obj2} gas cell now contains a large number of cells, which increases its complexity, and consequently, the time it takes to compute propagation of particles. However, it may still prove useful when in use with the extract function (see section \ref{Re-process legacy files}). Consider the following code:
\begin{verbatim}
    # Extract cells 5-10 from obj1
    objext = obj1.Extract(range(5,11))
    
    # Assemble two files by inserting cells
    obj2.InsertCells(objext)
\end{verbatim}
We first extract cells 5 to 10 from \texttt{obj1}, the resulting object, \texttt{objext}, now possesses an inadequate bounding surface (by default, a sphere of radius 20m). It is unsuitable to be inserted by copying the bounding surface, using the \texttt{Insert} function. However, it is still possible to assemble files with \texttt{InsertCells}, by excluding cells of \texttt{obj1} from \texttt{obj2}, allowing us to quickly assemble files without reworking the bounding surface.

Now, one could see that a problem arises when combining the two insert functions:
\begin{verbatim}
    obj2.Insert(obj1) # First insert by copying the bounding surface
    obj3.InsertCells(obj2) # Then insert by excluding cells from the gas cell
\end{verbatim}
Indeed, here, the gas cell of \texttt{obj1} is excluded from the gas cell of \texttt{obj3}. While this is still a correct input for MCNP, the resulting complexity of \texttt{obj3} gas cell is increased. Thus, one should always favor the use of \texttt{Insert} before \texttt{InsertCells}.

\subsection{Transform your file}

Before assembling two files, it is sometimes necessary to apply a MCNP transformation to properly position one file relative to the other. In MCNP-GO, this can be done with various functions. Users should refer to the online documentation for details, but the most useful are:	
\begin{itemize}
    \item \texttt{Transform}: input is the MCNP card formatted as a Python list. This is the most general way of doing a transformation, often useful when you already know the rotation matrix.
    \item \texttt{TrRotX}, \texttt{TrRotY}, \texttt{TrRotZ}: shift and rotates around axis X/Y/Z.
    \item \texttt{TrRotU}: shift and rotates around an arbitrary axis \texttt{u}.
\end{itemize}

Now, consider the following code:
\begin{verbatim}
    # Load MCNP files
    obj1 = go("file1.mcnp")
    obj2 = go("file2.mcnp")
        
    obj1.TrRotZ(angle=45) # 45° rotation of axis Z
    obj2.Insert(obj1) # Insert obj1 in obj2
    obj2.TrRotZ(angle=45) # 45° rotation of axis Z
\end{verbatim}
In this example, \texttt{obj1}, now inside \texttt{obj2}, has rotated by 90 degrees. Transformations apply to the whole file and not just the last inserted object. This logic follows what you would expect when assembling equipment in a real experiment. Note that it is possible to retrieve transformations applied to a object and then apply it to an other object, for duplication purposes, for instance:
\begin{verbatim}
    ... # Transforms were applied to obj2
    obj1.Transform(obj2.GetTr()) # Get that transformation and apply it to obj1
\end{verbatim}
In this example, \texttt{obj1} was transformed with transformations previously applied to \texttt{obj2}. This feature is also useful for defining tallies, as we shall see in section \ref{Keeping track of file properties}.
	
In practice, the transformation is applied to the object surfaces by updating the existing transforms cards or by adding one if the file does not possess one. Existing transformation cards are updated according to:
\begin{align}
M_{\text{new}} &= M_{\text{old}}M_{\text{in}} \label{equ:Msurf}\\
T_{\text{new}} &= M_{\text{in}}^tT_{\text{old}} + T_{\text{in}} \label{equ:Tsurf}
\end{align}
where $M$ and $T$ are respectively the rotation matrix and vector part of the transformation, subscript $(\text{in})$ refers to the input transform to be applied, $(\text{old})$ refers to the transformation card to be updated and $(\text{new})$ refers to the new (updated) transformation card.
	
Formulas (\ref{equ:Msurf}) and (\ref{equ:Tsurf}) do not apply to cell transformations (related to keywords \texttt{trcl} and \texttt{fill}) and are thus treated separately. With the same notations, cell transformations are updated according to:
\begin{align}
M_{\text{new}} &= M_{\text{in}}^tM_{\text{old}}M_{\text{in}}\\
T_{\text{new}} &= M_{\text{in}}^tT_{\text{old}} - M_{\text{new}}^tT_{\text{in}} + T_{\text{in}}
\end{align}
Note that since the treatment for surface and cell transformations are different, a new transformation card must be created for the latter.
	
For simplicity, the \texttt{*tr} cards are converted to \texttt{tr} cards (from degrees to cosinuses), and reverse transformations are converted to forward transformations. Moreover, in order to insure that rotated objects that are in contact (i.e. share a common surface), but do not possess a common transformation card, do not collide when the assembly is rotated, transformations are formatted with 15 digits precision. In particular, this problem arises when cell transformations and surface transformations coexist in the same file. Since update formulas, described above, are not the same, they do not share a common transformation card. Then, there is a possibility of collision due to floating number precision issues. However, if that problem arises, it is always possible to crop one of the cell with the bounding surface of the other. Note that this issue would not arise with cells that belong to a universe since they would be cropped by the universe's bounding surface.
	
An other way to apply a transformation to a file would be to transform surfaces parameters (as MCNP does at running time). This was not done for two reasons: (i) the output file would be less readable, (ii) this could lead to cell collisions within a file because of floating number precision issues.

\subsection{Keeping track of file properties: tallies and other features}
\label{Keeping track of file properties}
In order to define a source or a tally, one may need to keep track of cell, surface or transformation cards that are renumbered in the assembling process. To this end, it is possible to store such information in a json string, placed after the input file card block. This section of the input file is not read by MCNP and is usually used to store various information. Now, consider the following json string example:
\begin{verbatim}
{"ScintillatorCell":{"cell":[1],"surf":[1],"trans":[2],
"position":[0.0,1.0,0.0],
"comment":"Scintillator cell of detector 1"}}
\end{verbatim}
It contains information about the scintillator structure and position. The keys "cell", "surf" and "trans" contains list of cell, surface and transform numbers and should not contain other information as they are renumbered during assembly. These numbers can then be retrieved with a special function and used to build any card. Consider this simplest example of building a track length tally with the scintillator body:
\begin{verbatim}
    # You have successfully assembled your files and reached this point
    
    # Retrieve scintillator cells
    tiCells = obj.GetGroup("ScintillatorCell","cell")
    sComment = obj.GetGroup("ScintillatorCell","comment")
    
    # We build a tally card and add it
    lsTallyCard = [sComment, f"F4:P {tiCells[0]}"]
    obj.AddMCNPCard(lsTallyCard)
\end{verbatim}
With this method, one could define tailored functions to build any MCNP card based on information stored in the json string. There is no limit to the numbers of entries that it can contains but there must be only one json string. Other key entries are also possible, only "cell", "surf", "trans" and "comment" keys are reserved. 

In particular, one could retrieve the position of a point detector or of a source and apply a transform to it. In the above json string, the key "position" refers to the point detector position located at the center of the scintillator. The following code shows how such procedure could be done: 
\begin{verbatim}
    # You have successfully assembled your files and reached this point
    
    import numpy as np
    
    # Retrieve position of point detector and transform number
    npPos = np.array(obj.GetGroup("ScintillatorCell","position"))
    iTr = obj.GetGroup("ScintillatorCell","trans")[0]
    
    # Read transform card in file
    dictTrCard = obj.FindTrCard(iTr)
    npTransCard = np.array(dictTrCard["translat"])
    npMatCard = np.array(dictTrCard["rot"]).reshape(3,3)
    
    # Detector position in final assembly
    npFinalPos = npMatCard.T.dot(npPos) + npTransCard
\end{verbatim}
In this example, imagine that the scintillator was inserted in other files and many transformations were applied through the course of your assembly. The position and orientation of the scintillator can be back traced using the json scintillator transform number (here, it was originally number 2). This transformation number was renumbered during assembly but the new number can be retrieved from the json string and with it the transformation card. The updated position $P_{\text{new}}$ is then: $P_{\text{new}} = M^tP_{\text{old}} + T$, with $P_{\text{old}}$ the old position, $M$ and $T$ respectively the matrix and vector part of the transformation.

\subsection{Traceability}
\label{Traceability}
	
During assembly, MCNP-GO keeps track of the operations that were performed on your file	and the objects contained in it. Transformations and insertions, together with files paths are specified in the header part of the file. Consider the following header extracted from next section example:
\begin{verbatim}
c  - Original file: 
c ./room.mcnp
c      No transforms were applied
c  - Inserted files: 
c ./detector.mcnp
c      Applied translation: [0.0, 400.0, 0.0]
c      Applied Euler XZX angles: a=-90.0, b=1.0, g=90.0 
c      Rotation matrix:
c           [ 0.9998477   0.         -0.01745241]
c           [0. 1. 0.]
c           [0.01745241 0.         0.9998477 ]
c      List of applied transforms:
c           Translation: [0, 400, 0] Rotation Y: 1
c       - Files contained in ./detector.mcnp :
c       ./ccd.mcnp
c            Applied translation: [60, 50, 0]
c            Applied Euler XZX angles: a=-0.0, b=0.0, g=0.0 
c            Rotation matrix:
c                 [1 0 0]
c                 [0 1 0]
c                 [0 0 1]
c            List of applied transforms:
c                 Translation of vector: [60, 50, 0]
c ./lat_ex5.mcnp
c      Applied translation: [0.0, 300.0, 0.0]
c      Applied Euler XZX angles: a=45.0, b=0.0, g=0.0 
c      Rotation matrix:
c           [0.70710678 0.70710678 0.        ]
c           [-0.70710678  0.70710678  0.        ]
c           [0. 0. 1.]
c      List of applied transforms:
c           Translation: [0, 300, 0] Rotation Z: 45
\end{verbatim}
The header structure is recursive, and allows to reconstruct every steps that was taken to build this file. This feature is in line with the systems engineering approach of the package. Moreover, if the files are managed using configuration management system, then every modifications related to file geometry or positioning can be back traced to its source. This could be very valuable for applications with very complicated input files when it comes to comparing different designs.

\section{Example use case}
\label{Example use case}	

In this section, we show an example of how MCNP-GO could be used to assemble the MCNP input file of a tomographic experiment. In such experiments, an object of interest is radiographed with an X-ray source. A large collection of radiographies, each corresponding to a angle of rotation of the object, is then necessary to reconstruct the object's interior. To perform high precision tomography, it is necessary to precisely know the source, object and detector position and orientation. Say that this is done with a laser tracker device, that gives position and orientation relative to the room reference frame, and you wish to build a numerical twin of your experiment. Then, MCNP-GO is the right tool for that. In our example, the steps taken to assemble the MCNP input file reflect what is done in the real experiment.

\begin{figure}[h!]
\includegraphics[scale=0.1]{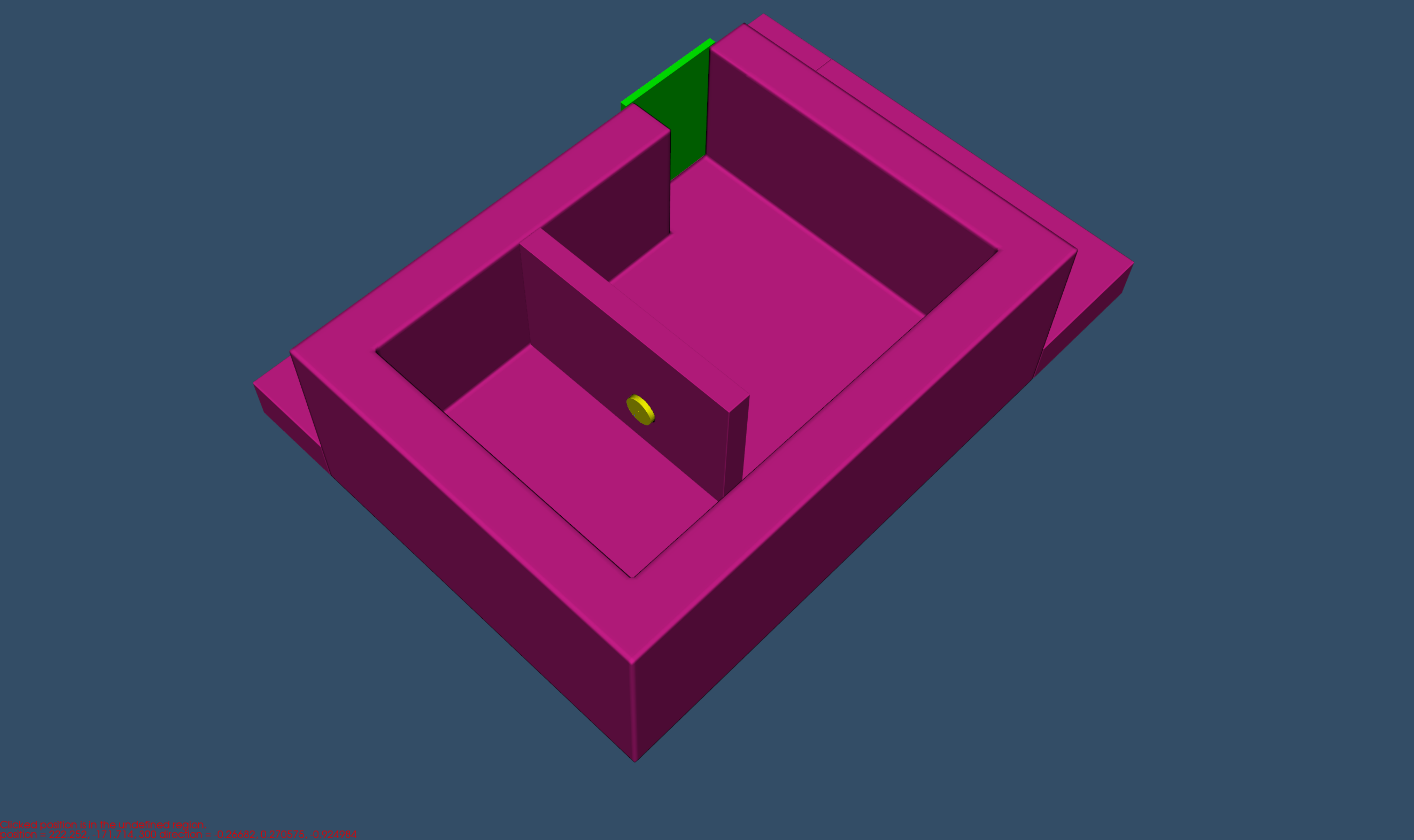}
\caption{MCNP model of the experience room showing concrete walls and floor (purple), lead door (green), and primary tungsten beam collimation (yellow). The tantalum target is too small to be seen.\label{fig:room}}
\end{figure}

\begin{figure}[h!]
\includegraphics[scale=0.05]{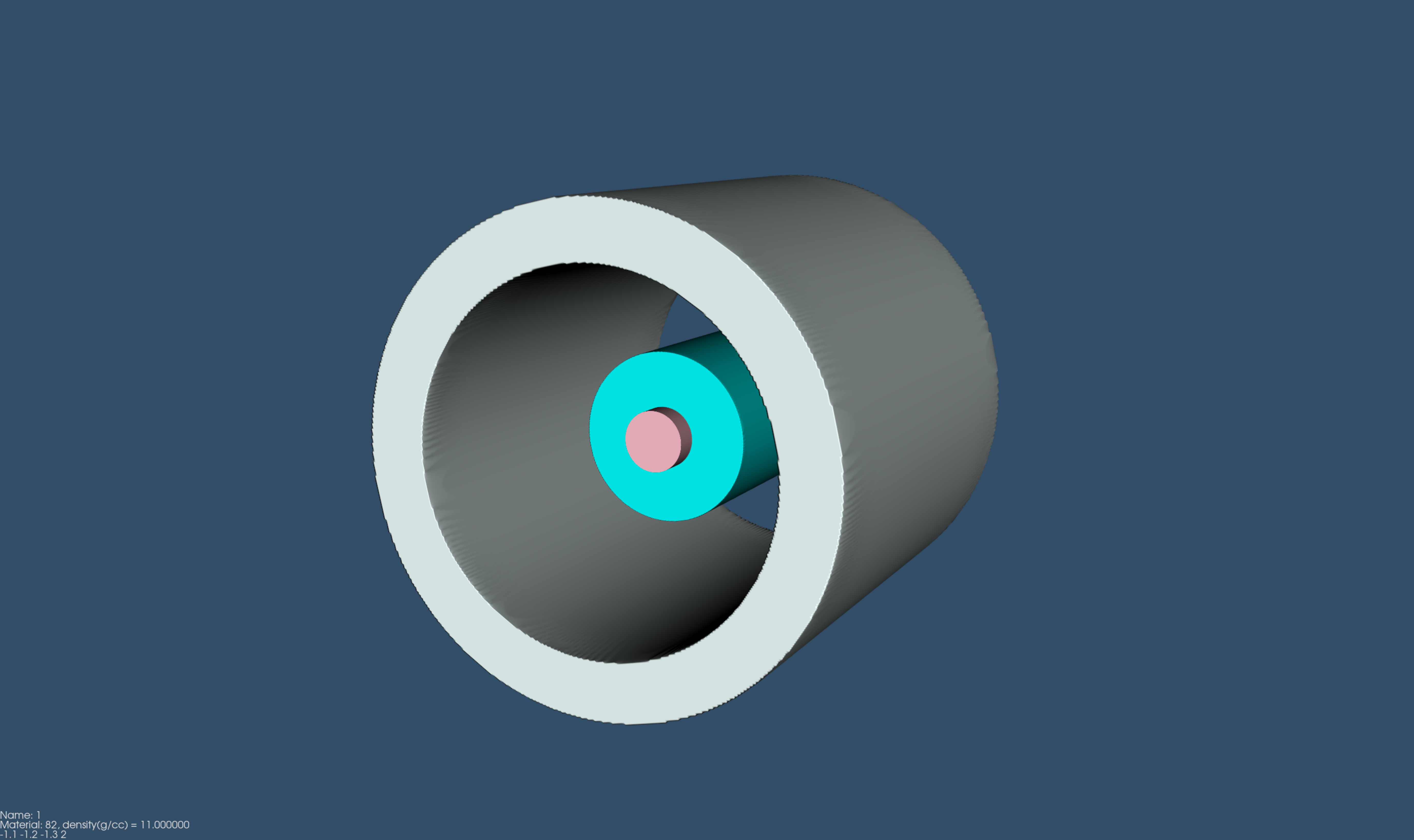}
\caption{MCNP model of the CCD camera, placed inside a lead cylinder to shield from radiations.\label{fig:ccd}}
\end{figure}

\begin{figure}[h!]
\includegraphics[scale=0.1]{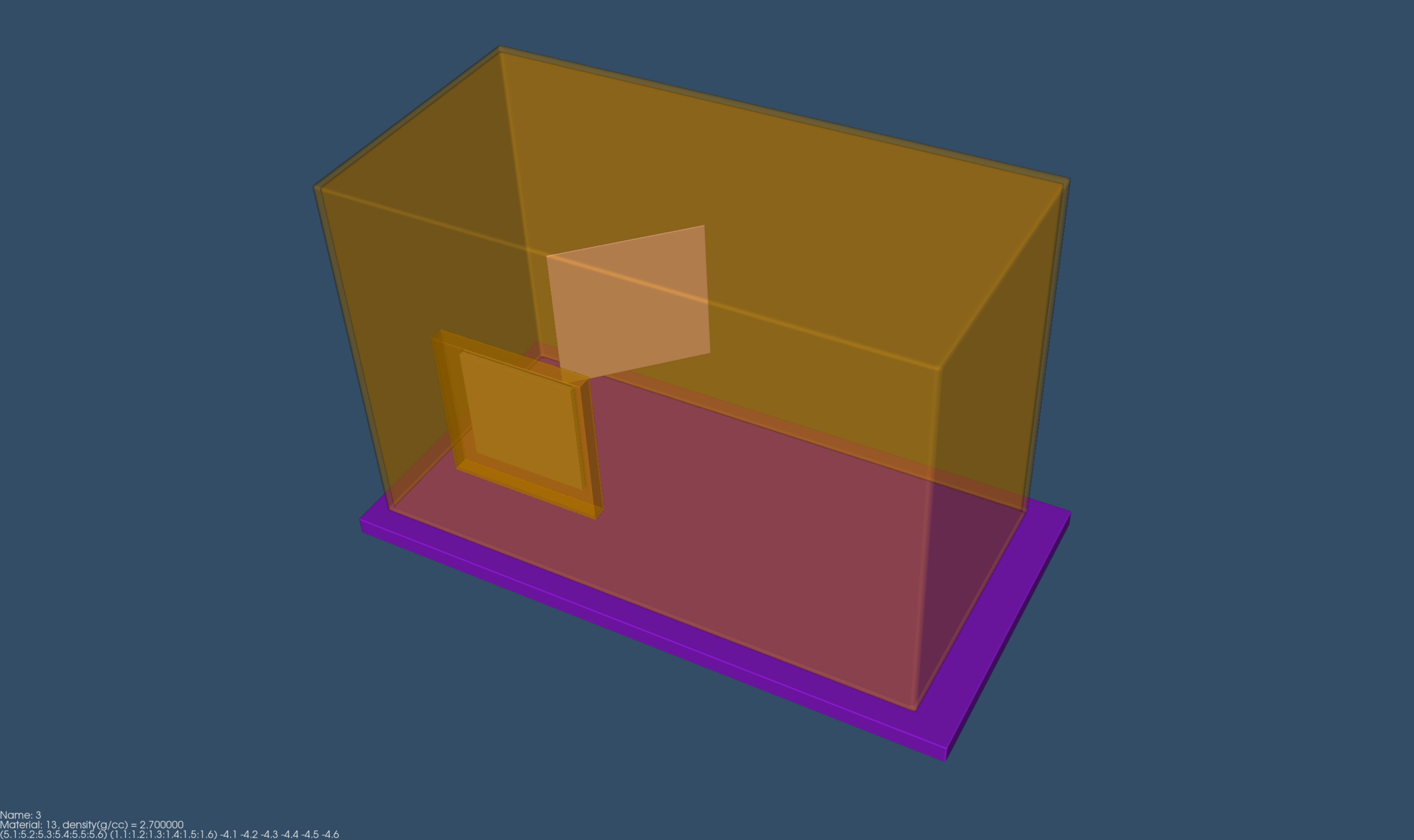}
\caption{MCNP model of the detector bench, containing a mirror and a scintillator on the front side.\label{fig:detector}}
\end{figure}

\begin{figure}[h!]
\includegraphics[scale=0.2]{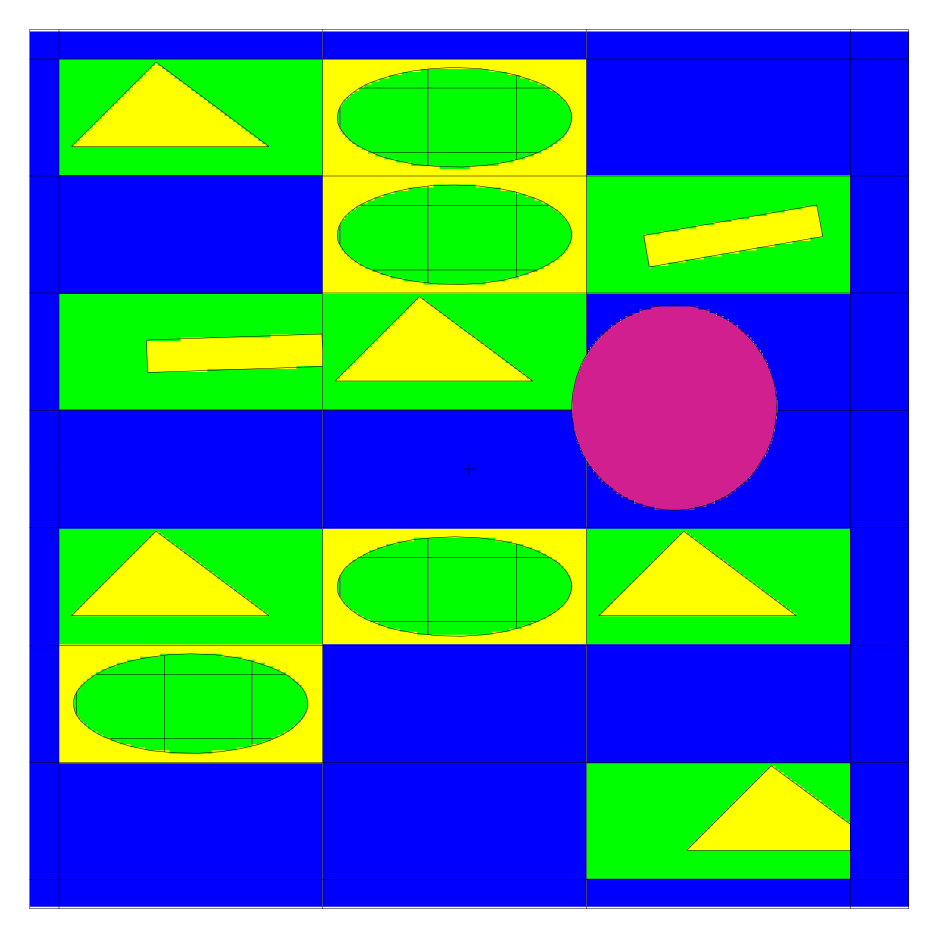}
\caption{MCNP cut plot of the test object (lattice example number 5 of the MCNP manual). \label{fig:lat5}}
\end{figure}

Our geometry is made of the following objects:
\begin{itemize}
    \item An experience room, containing an X-ray source. In this example, X-rays are produced by focusing an electron beam toward a tantalum target. A tungsten collimation is placed in front of the target to shape the X-rays into a beam (aligned with axis $Y$).
    \item A detector bench, containing a Bismuth Germanate Oxyde (BGO) scintillator on the front side and a mirror to reflect photons toward a CCD camera.
    \item A CCD camera, to be inserted inside the detector bench.
    \item A test object to be characterized. For this object, in order to illustrate MCNP-GO's capabilities, we choose lattice example number 5 from the MCNP manual. It was slightly reworked to obey rules described in section \ref{Requirements and caveats}, i.e. the gas and graveyard cells were placed at the end of the cell block.
\end{itemize}
Models are shown in Figures \ref{fig:room} to \ref{fig:lat5}. The 3D renderings were done with the GXSVIEW software \cite{GXSVIEW}. In order to assemble our input file, we need to perform the following steps:
\begin{itemize}
\item Step 1: Shift the CCD camera and insert it inside the detector bench.
\item Step 2: Shift and rotate the detector bench (now containing the CCD) and insert it inside the experience room. For illustration purposes, the detector is rotated by 1 degree around axis $Y$.
\item Step 3: Shift and rotate the test object and insert it inside the experience room.
\item Step 4: Save input file.
\item Step 5: Repeat steps 3 and 4 for each rotation angle of the test object.
\end{itemize}

\begin{figure}
\includegraphics[scale=0.1]{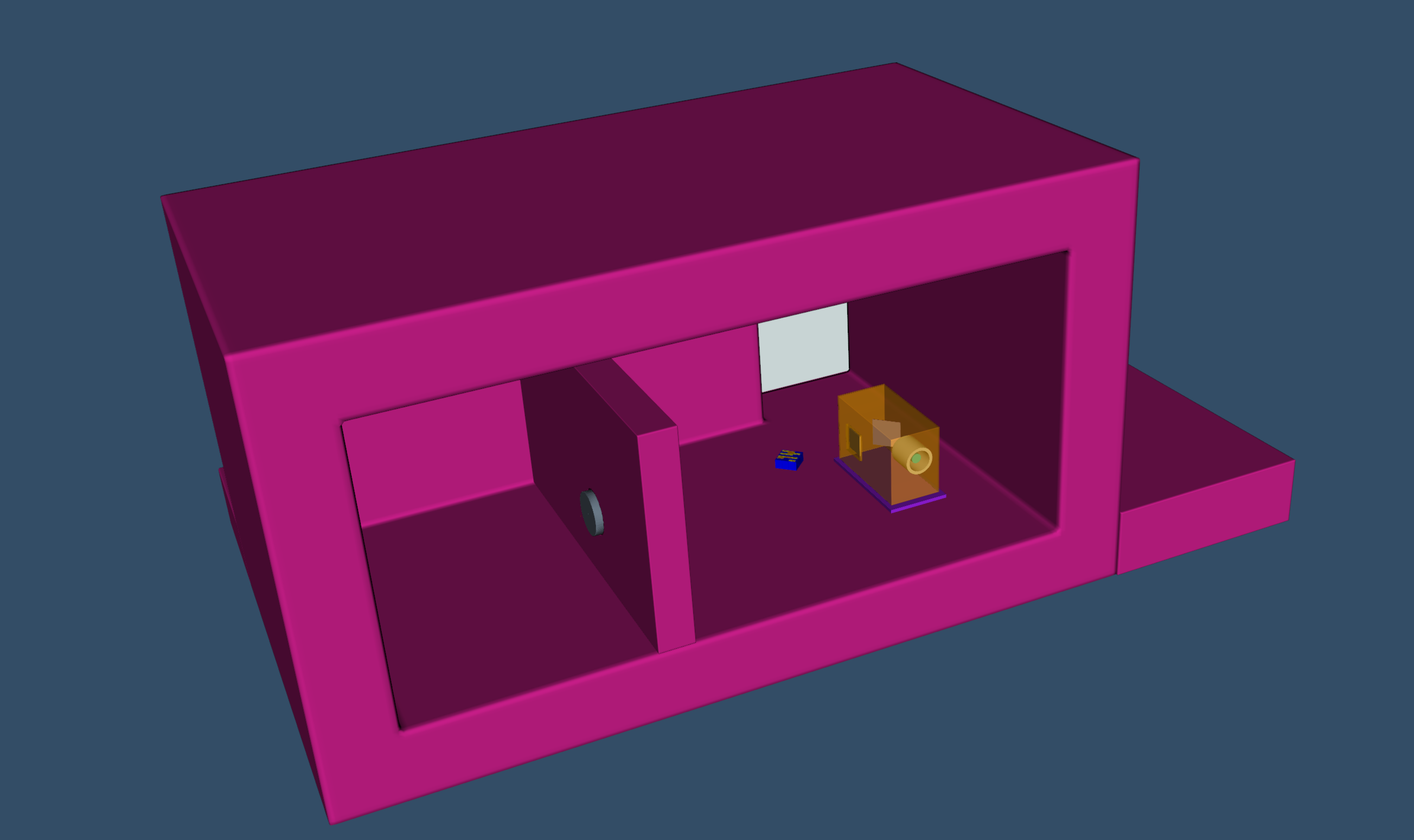}
\caption{MCNP model of the final assembly. The detector bench, CCD camera and lattice test object were inserted and placed in the experience room. \label{fig:room_45}}
\end{figure}

The Python script that replicates our experiment is then:
\begin{verbatim}
from mcnpgo.mcnpgo import *
from copy import deepcopy

# Loading files
room = go("./room.mcnp")         # experience room
detector = go("./detector.mcnp") # detector bench
ccd = go("./ccd.mcnp")           # CCD camera
lat = go("./lat_ex5.mcnp")       # test object

# Start by placing the CCD in the detector bench
ccd.Translat([60,50,0])
detector.Insert(ccd, location = 'inside')

# Moving detector+ccd
# Detector bench is tilted by 1° around axis Y
detector.TrRotY(trans=[0,400,0],angle=1)

# Insert detector in experience room
room.Insert(detector, location = 'inside')

# Loop over test object angles
angles = [0,30,45,90]
for d in angles:
# Local copy of lattice and room
lat_cpy = deepcopy(lat)
room_cpy = deepcopy(room)

# Shift and rotate
lat_cpy.TrRotZ(trans=[0,300,0],angle=d)

# Insert in room
room_cpy.Insert(lat_cpy, location = 'inside')

# Save files
room_cpy.WriteMCNPFile(f"room_{d}.mcnp")
\end{verbatim}
An input file of the full assembly is generated for each angle of rotation. It is shown in Figure \ref{fig:room_45} with the test object rotated by 45 degrees. In this example, data cards other than geometry related (source, tally, etc...) were not added for the sake of concision. However, functions to add tallies and to help with the tasks of adding data cards are available, see online documentation for more details.

\begin{figure}
\includegraphics[scale=0.1]{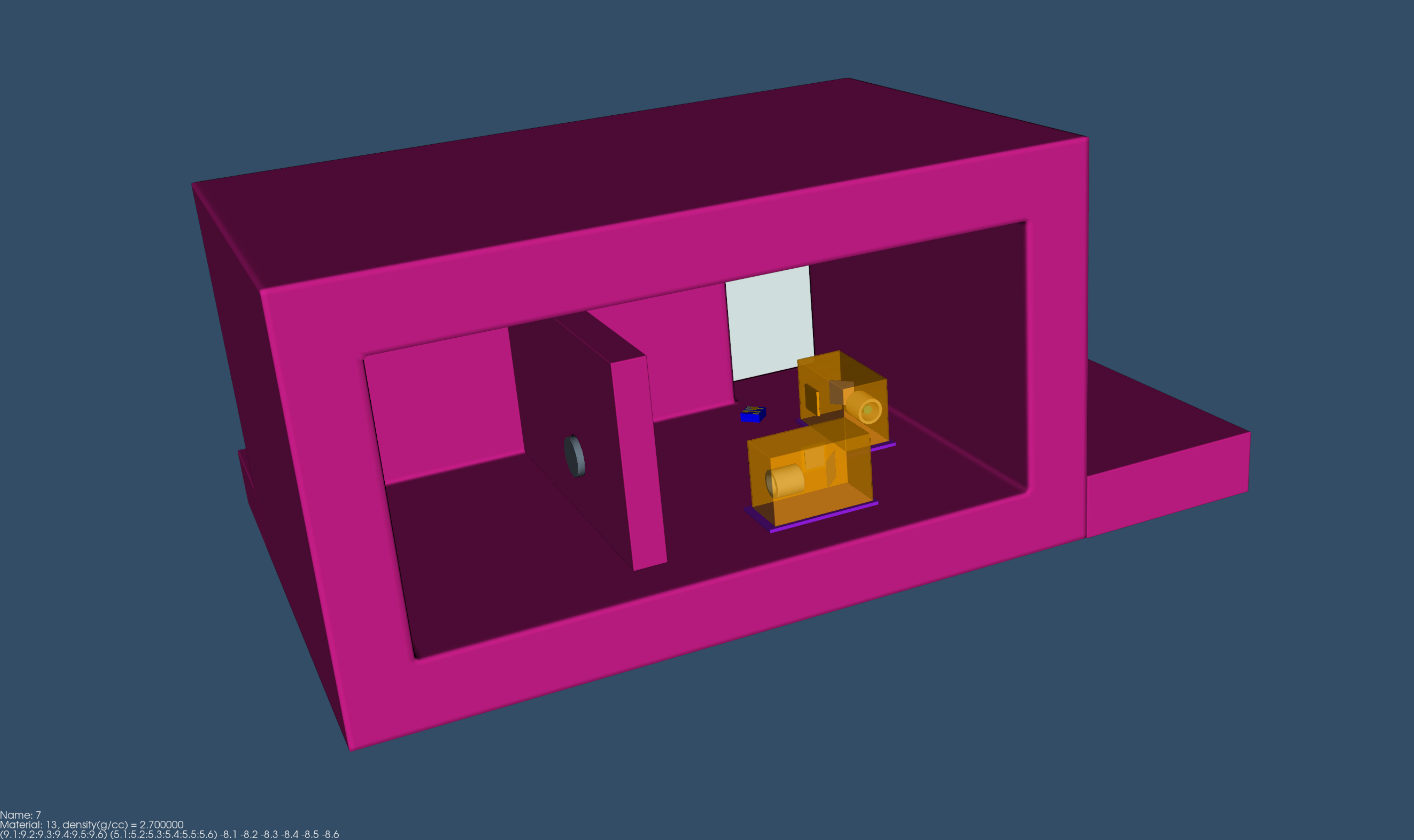}
\caption{MCNP model of the final assembly with the added detector bench.\label{fig:newroom_45}}
\end{figure}

Now, say that this Python script is lost and that someone wants to add a detector bench to the set up. All we have are the MCNP input files of the final assembly. With the help of the extract function, this can be easily done. We need to perform the following steps:
\begin{itemize}
\item Step 1: Extract detector bench from input file.
\item Step 2: Shift and rotate the detector to its original position. This step requires to know what transformations were applied to the detector bench, which is indicated in the header of the file (see section \ref{Traceability}).
\item Step 3: Shift and rotate to the desired position and insert it in the experience room.
\item Step 4: Save the file.
\end{itemize}
The corresponding Python script is then:
\begin{verbatim}
from mcnpgo.mcnpgo import *
from copy import deepcopy

# Load old file
room_45 = go("./room_45.mcnp") 

# Extract detector bench
detector = room_45.Extract(range(12,22))

# Reverse it back to its original position
detector.Translat([0,-400,0]) # Pure translation
detector.TrRotY(angle=-1)     # Pure rotation

# Place it inside experience room
detector.TrRotZ(angle=-90,trans=[200,300,0])

# Insert detector in experience room
# We need to use InsertCells since detector was extracted
room_45.InsertCells(detector)

# Save file
room_45.WriteMCNPFile("newroom_45.mcnp")
\end{verbatim}
The result is shown in Figure \ref{fig:newroom_45}. In a few steps the new detector bench was duplicated and rotated by a 90 degree angle.


\section{Conclusions and future work}
\label{Conclusions and future work}

In this article, MCNP-GO, a python package to assemble MCNP input files was presented. While knowledge of the most basic python features is necessary, it was designed with non python users in mind. The guiding principle is to consider each MCNP input file as an independent object, the nature of which is not important, but it must be a valid MCNP file. These objects form a database from which the user can pick to assemble an input file. The python script of the assembly then reflects the logic of the real experiment, making it intuitive for users. MCNP-GO is in use at CEA and has tremendously simplified the work flow of generating input files. The package is not specifically designed to generate MCNP source, tally or data cards but still provide basic tools for it. It is then up to users to create tailored functions for their needs. 

MCNP-GO was built with a systems engineering approach in mind. Meaning that each object can be considered as a system with its own reference frame. If the user's database is managed with configuration management system, then the output file final assembly contains every information necessary to rebuild it and each modification can be back traced to its source. This ensures reliability and traceability of the work flow.

Future works will focus on different methods for assembling files. We saw in section \ref{Assembling files} two methods for assembling files: one makes use of the boundary surfaces and the other of the exclusion operator. However, many more methods could be imagined depending on user needs. For our purposes, using a unique gas cell results in a reasonable complexity, but this would not be the case for very complex parts used in nuclear fusion energy simulations (see \citep{ALGUACIL2018} for a discussion on this subject). Indeed, nuclear fusion MCNP models are generated using CAD to MCNP conversion algorithms \cite{WU2009,GEOUNED} resulting in high fidelity MCNP models, at the cost of cell complexity. In such circumstances, it would be unreasonable to define a unique gas cell, the cell complexity would be too high. Instead, it is split in simpler pieces by a void generation algorithm \cite{GEOUNED_void}. In this case, insert methods described in section \ref{Assembling files}, relying on a unique gas cell, would not work. To solve this problem, one would need to extract the relevant cells from each file, assemble them, and split the gas cells with a void generation algorithm at the very end.

\section*{Conflict of Interest Statement}
The author declares that the research was conducted in the absence of any commercial or financial relationship that could be construed as a potential conflict of interest.	
	
\section*{Funding}	
No funding information applicable.

\bibliography{document}

\end{document}